\documentstyle[12pt]{article}
\title{Analysis   of   the  pion  wave  function   in   light-cone
formalism\thanks{Project  supported by National  Natural  Science
foundation of China and the grant LWTZ-1298 of Chinese Academy of
Science.}}
\author{Tao Huang, Bo-Qiang Ma and Qi-Xing Shen \\
        Center  of Theoretical Physics, China Center of  Advanced
        Science \\
        and Technology (world Laboratory) Beijing, China \\
        and \\
        Institute of High Energy Physics,
        Academia Sinica\\
        P.O.Box  918(4),  Beijing  100039,   China\thanks{Mailing
        Address}}
\date{}
\begin{document}
\maketitle

\baselineskip 20pt
\begin{abstract}
Several general constraints are suggested to analyze the pionic
valence-state wave function. It is found that the present model
wave functions used in light-cone formalism of perturbative
quantum chromodynamics have failed these requirements to fit  the
pionic formfactor data and the reasonable valence-state structure
function which does not exceed the pionic structure function data
for $x \rightarrow 1$ simultaneously. Furthermore, it is  pointed
out  that  there is a possibility to find  model  wave  functions
which  can satisfy all of general constraints. Also we show  that
there  are two higher helicity $(\lambda_{1} + \lambda_{2} =  \pm
1)$ components in the light-cone wave function for the pion as  a
natural consequence from the Melosh rotation and it is speculated
that   these   components  should  be   incorporated   into   the
perturbative quantum chromodynamics.
\end{abstract}

\vspace{5mm}
Detailed Version. \\To be published in Phys.Rev.D
\newpage
\section*{I. Introduction}

{\hskip  0.6cm}The hadronic wave functions in terms of quark  and
gluon  degrees of freedom play an important role in  the  quantum
chromodynamics  (QCD) predictions for hadronic processes. In  the
perturbative  QCD  theory  (pQCD)$^{1}$,  the  hadronic distribution
amplitudes  and  structure functions which  enter  exclusive  and
inclusive  processes  via the factorization  theorems$^{2-3}$  at
high  momentum  transfer can be determined by the  hadronic  wave
functions,  and therefore they are the underlying  links  between
hadronic  phenomena in QCD at large distances  (non-perturbative)
and   small  distances  (perturbative).  If  the  hadronic   wave
functions  were  accurately known, then we  could  calculate  the
hadronic  distribution  amplitudes and  structure  functions  for
exclusive  and  inclusive  processes in  QCD.  Conversely,  these
processes  also can provide phenomenological constraints  on  the
hadronic   distribution   amplitudes,  the   hadronic   structure
functions, and thereby the hadronic wave functions.

\vspace{0.5cm}

Several  important  non-perturbative tools  have  been  developed
which  allow specific predictions for the  hadronic  distribution
amplitudes  or the hadronic wave functions directly from  theory.
QCD  sum-rule technique$^{4-5}$ and lattice gauge  theory$^{6-7}$
provide  constraints on the moments of the hadronic  distribution
amplitude.  One  thus  could  model  the  hadronic   distribution
amplitudes  by  fitting  the  first  few  moments  in  terms   of
Gegenbauer  polynomials which are the solutions to the  evolution
equation   of   hadronic  distribution   amplitude$^{2-3}$.   The
Chernyak-Zhitnitsky (CZ) distribution amplitudes$^{4-5}$  (dubbed
wave  functions in the original papers) constructed in  this  way
are good in reproducing the correct sign and magnitude as well as
scaling  behavior of the pion, proton and neutron  form  factors.
However,  whether  the  CZ-like  distribution  amplitude  is  the
correct pion distribution amplitude is still an open problem  and
its  correctness  should  not be judged by only  its  success  in
reproducing  the correct magnitude of the pion form factor.  Some
earlier  lattice  Monte  Carlo  calculations$^{6}$,  designed  to
compute the pion distribution amplitude directly, were unable  to
distinguish  between  the asymptotic form and the CZ form.  In  a
recent  improved lattice QCD calculation$^{7}$ the second  moment
of  the  pion distribution amplitude was found to be  small  than
previous    lattice   calculations$^{6}$   and    the    sum-rule
calculations$^{4-5}$,   and   this   suggests   that   the   pion
distribution  amplitude may close to the asymptotic form  rather
than  the CZ form. From another point of view, as different  wave
functions may give a same distribution amplitude, there are still
ambiguities  about  the wave function even if we know  the  exact
form  of the distribution amplitude. Hence it is still  necessary
to  develop methods which specify the hadronic  wave  functions
directly.

\vspace{0.5cm}

In  principle,  the Bethe-Salpeter formalism$^{8}$ and  the  more
recent  discretized light-cone quantization approach$^{9}$  could
determine the hadronic wave functions, but in practice there  are
many  difficulties  in  getting  the  exact  wave  functions   at
present$^{10-11}$. One useful way is to use the approximate bound
state  solution  of a hadron in terms of the quark model  as  the
starting  point for modeling the hadronic valence wave  function.
The Brodsky-Huang-Lepage (BHL) prescription$^{3}$ of the hadronic
wave  function is in fact obtained in this way by connecting  the
equal-time wave function in the rest frame and the wave  function
in  the  infinite  momentum frame,  and  the  corresponding  wave
function  gives  a distribution amplitude  which  is  significant
different  from  the  CZ-form.  In  order  to  give  the  CZ-like
distribution amplitude which is required to fit the  experimental
data,  a  phenomenological model for the hadronic  valence  state
wave function has been proposed$^{12-14}$ by adding a  factorized
function to the BHL wave function. The distribution amplitude  is
almost  the same as the CZ distribution amplitude except for  the
end-point regions. Recently a light-cone quark model approach  of
hadrons$^{15}$  has  received attention for the reason  that  the
model  can simultaneously fit low energy phenomena, the  measured
high   momentum   transfer  hadron  form  factor,  and   the   CZ
distribution  amplitudes.  The  hadronic wave  function  in  this
approach,  as  will be shown, is significant different  from  the
factorized wave functions in Refs.12-13 though both of them  give
the similar CZ-like distribution amplitudes.

\vspace{0.5cm}

The  purpose  of this paper is to discuss  the  consequence  from
physical  constraints on the hadronic wave functions, using  pion
valence  state  wave function as an example. In Sec.II,  we  will
list  several general constraints on the pion valence state  wave
function$^{3,16}$. We then analyse, in Sec.III, several  existing
pion wave functions used in light-cone formalism of  perturbative
QCD.  We hope to find a wave function which could give  both  the
approximate  CZ distribution amplitude required to fit  the  pion
form  factor  data  and the reasonable  valence  state  structure
function  which does not exceed the pion structure function  data
simultaneously,  with  several  constraints  on  the  pion   wave
functions  also  satisfied.  We  find,  unfortunately,  that  the
present  model  wave  functions  have  failed  the  requirements.
Furthermore, the most recent light-cone quark model wave function
given  in Ref.15 violates the general constraints  severely.  For
example, it gives a probability of finding the valence Fock state
in a pion much larger than unity if the correct normalization  of
the  pion  distribution amplitude is retained.  Hence  this  wave
function  suffers  from serious flaws. However we show  that  the
power-law  form  of  the pionic wave  function  can  satisfy  the
requirements  and it provides a possible example to find  a  good
wave  function  to be used in QCD theory. In  order  to  properly
evaluate the effect from Melosh rotation$^{17-18}$ connecting the
rest  frame  equal-time wave function and  the  light-front  wave
function  in the light-cone formalism, we re-construct in  Sec.IV
the  light-cone quark model of the pion in light-front  dynamics.
We  find  that  the contributions from  Melosh  rotation  in  the
ordinary helicity $(\lambda_{1} + \lambda_{2} = 0)$ component
wave  function  by using reasonable parameters seem  to  have  no
significant  effect on the calculated distribution amplitude  and
structure  function  in comparison with those from the  BHL  wave
function.   However,   two  higher   helicity   $(\lambda_{1}   +
\lambda_{2}  =  \pm  1)$ components which  were  not  taken  into
account in previous perturbative QCD emerge naturally in the full
light-cone  wave  function  as a  consequence  from  the  Wignier
rotation. In Sec.V we present the summary and some comments.

\section*{II. Constraints on the Valence-state Wave Function}

{\hskip   0.6cm}In  this  paper,  we  employ   the   particularly
convenient light-cone formalism$^{1-3}$ in which the  description
of  the hadronic wave functions is given by a set  of  n-particle
momentum space amplitude,
$$
\psi_{n}(x_{i},  \vec{k}_{\bot i}, \lambda_{i}), ~ i  =  1,2,...n
\eqno (2.1)
$$
defined  on the free quark and gluon Fock basis at equal  "light-
cone time" $\tau = t + z$ in physical "light-cone" gauge $A^{+} =
A^{0}  +  A^{3}  =  0$.  Here  $x_{i}  =  k^{+}_{i}/p^{+}$,  with
$\sum_{i}  ~x_{i}  = 1$, is the light-cone momentum  fraction  of
quark  or gluon $i$ in the n-particle Fock state;  $\vec{k}_{\bot
i}$,  with  $\sum_{i} ~\vec{k}_{\bot i} = 0$, is  its  transverse
momentum   relative   to  the  total  momentum   $p^{\mu}$;   and
$\lambda_{i}$ is its light-cone helicity. Any hadron state can be
expanded  in terms of this complete set of Fock states  at  equal
$\tau$:
$$
\begin{array}{rl}
\mid  H > = \sum_{n,\lambda_{i}}  \int  [dx][d^{2}\vec{k}_{\bot}]
\psi_{n}(x_{i},    k_{\bot   i},   \lambda_{i})
&\prod_{fermions}
\frac{u(x_{i}   p^{+},  x_{i}  \vec{p}_{\bot}   +   \vec{k}_{\bot
i})_{\lambda_{i}}}{\sqrt{x_{i}}} \\
&\prod_{gluons} \frac{\epsilon(x_{i} p^{+}, x_{i}
\vec{p}_{\bot}  + \vec{k}_{\bot  i})_{\lambda_{i}}}{\sqrt{x_{i}}}
\mid n > \end{array}  \eqno (2.2)
$$
with the normalization condition,
$$
\sum_{n,    \lambda}    \int    [dx][d^{2}\vec{k}_{\bot}]    \mid
\psi_{n}(x_{i}, \vec{k}_{\bot_{i}}, \lambda_{i}) \mid^{2} = 1  ~,
\eqno (2.3)
$$
where the sum is over all Fock states and helicities, and
\begin{eqnarray*}
[dx]  &  =  & \delta(1 -  \sum^{n}_{i=1}  x_{i})  \prod^{n}_{i=1}
{}~dx_{i} ~~, \\
{[d^{2}\vec{k}_{\bot}]} &=& 16 ~\pi^{3} ~\delta^{2} (\sum^{n}_{i=1}
{}~\vec{k}_{\bot   i})   \prod^{n}_{i=1}   (d^{2}    ~\vec{k}_{\bot
i}/16\pi^{3})~~.
\end{eqnarray*}
The  quark  and  gluon structure functions  $G_{q/H}(x,  Q)$  and
$G_{g/H}(x,Q)$,  which control hard inclusive reactions, and  the
hadron  distribution amplitudes $\phi_{H}(x, Q)$,  which  control
hard  exclusive  reactions,  are simply  related  to  these  wave
functions$^{1-3}$:
$$
G_{a/H}(x,  Q) = \sum_{n} \int [d^{2} ~\vec{k}_{\bot  i}][dx_{i}]
\mid \psi_{n}(x_{i}, \vec{k}_{\bot i}) \mid^{2}  \delta(x-x_{a}),
{}~a=q ~~or ~~g~; \eqno (2.4)
$$
and
$$
\phi_{H}(x,   Q)   =  \int^{Q^{2}}  [d^{2}   ~\vec{k}_{\bot   i}]
\psi_{valence} (x_{i}, \vec{k}_{\bot i}) ~. \eqno (2.5)
$$
In the case of inclusive reactions all of the hadron Fock  states
generally participate; whereas in the case of exclusive reactions
perturbative  QCD predicts that only the lowest  particle  number
(valence) Fock state contributes to the leading order in $1/Q$.

\vspace{0.5cm}

In principle the hadronic wave functions determine all properties
of  hadrons.  From the relation between the  wave  functions  and
measurable quantities we can get some constraints on the  general
properties of the hadronic wave functions. In the pionic case two
important constraints on the valence state wave function (for the
$\lambda_{1}   +   \lambda_{2}  =  0$   components)   have   been
derived$^{3}$  from  $\pi  \rightarrow  \mu  \nu$  and   $\pi^{0}
\rightarrow \gamma \gamma$ decay amplitudes:
$$
\int^{1}_{0}    dx   ~\int    (d^{2}    \vec{k}_{\bot}/16\pi^{3})
\psi_{q\bar{q}} (x, \vec{k}_{\bot}) = f_{\pi}/2\sqrt{3} ~~, \eqno
(2.6)
$$
and
$$
\int^{1}_{0}  dx  ~\psi_{q\bar{q}}  (x,  \vec{k}_{\bot}  =  0)  =
\sqrt{3}/f_{\pi} ~~, \eqno (2.7)
$$
where  $f_{\pi} \approx 93 ~MeV$ is the pion decay constant.  The
$\lambda_{1}  + \lambda_{2} \neq 0$ component valence state  wave
functions do not contribute in the $\pi \rightarrow \mu \nu$
and $\pi^{0} \rightarrow \gamma \gamma$ processes thereby   the
presence  of  the $\lambda_{1} + \lambda_{2} \neq  0$  components
does  not  alter the above two  constraints.  Experimentally  the
average    quark    transverse    momentum    of    the     pion,
$<\vec{k}^{2}_{\bot}>_{\pi}$,  is  of the order  $(300  MeV)^{2}$
approximately$^{16}$.  The  quark  transverse  momentum  of   the
valence state pion, defined to be
$$
<\vec{k}^{2}_{\bot}>_{q\bar{q}}        =       \int        (d^{2}
\vec{k}_{\bot}/16\pi^{3})  dx \mid \vec{k}^{2}_{\bot}  \mid  \mid
\psi_{q\bar{q}}(x,   \vec{k}_{\bot})  \mid^{2}/p_{q\bar{q}}   ~~,
\eqno (2.8)
$$
should  be  large than $< \vec{k}^{~2}_{\bot} >_{\pi}$.  We  thus
could require that $\sqrt{ < \vec{k}^{~2}_{\bot} >_{q\bar{q}}}$  have
the  value  of  about a few hundreds MeV, serving  as  the  third
constraint.  The fourth constraint is the most natural  one:  The
probability of finding the $q\bar{q}$ Fock state in a pion should
be not larger than unity,
$$
P_{q\bar{q}}  = \int (d^{2} \vec{k}_{\bot} / 16\pi^{3})  dx  \mid
\psi_{q\bar{q}}(x,  \vec{k}_{\bot})  \mid^{2} \leq  1  ~~.  \eqno
(2.9)
$$

For the distribution amplitude we compare the calculated one with
the CZ-form,
$$
\phi_{CZ}(x)  =  5\sqrt{3} ~f_{\pi} x(1-x)(2x -1)^{2}  ~~.  \eqno
(2.10)
$$
As  to the case of structure function, it should be noticed  that
the  valence  state structure function is only one  part  of  the
valence  structure function of a pion. Hence it is reasonable  to
require that the calculated valence state structure function  not
exceed the structure function data for the pion. We will use  the
NA3  parameterization$^{19}$  of the pion structure  function  in
comparison with the calculated valence state structure  function.
It should be indicated that the $Q^{2}$ corresponding to the  NA3
data$^{19}$  is  very large; i.e., $Q^{2} =  25(GeV/c)^{2}$.  For
$Q^{2}$  of  the  order  of a  few  $(GeV/c)^{2}$  the  structure
function  should increase at large x and decrease at small  x  in
considering    the    contributions    from    QCD    logarithmic
evolution$^{20}$,  the higher twist effects and  other  power-law
type  sources$^{21}$  for Bjorken scaling  violations,  with  the
shape and magnitude not changed too much.

\section*{III. Analysis of Several Existing Wave Functions}

{\hskip  0.6cm}The hadronic wave function depends essentially  on
the   non-perturbative  QCD theory,  and  it  exhibits  the   full
complexity  of non-perturbative dynamics. It is necessary for  us
to use both theoretical tools and phenomenological constraints in
studying  the hadronic wave functions. In order to get  clear  on
where  assumptions  and approximations have been made  and  where
problems  may occur, we need to review some previous results  in
the following analyses of several existing pionic wave functions.

\vspace{0.7cm}

\noindent
{\bf A. The Brodsky-Huang-Lepage Prescription}

Brodsky-Huang-Lepage  suggested$^{3}$  a connection  between  the
equal-time  wave  function in the rest frame and  the  light-cone
wave  function by equating the off-shell propagator  $\epsilon  =
M^{2} - ( \sum^{n}_{i=1} k_{i})^{2}$ in the two frames:
$$
\epsilon = \left \{ \begin{array}{l}
M^{2} - ( \sum^{n}_{i=1} ~q^{0}_{i} )^{2} ~, ~~~~~~~~~~~~~~~~~~~~
\sum^{n}_{i=1} ~\vec{q}_{i} = 0 ~~~~{[C.M.]} \\
M^{2}    -    \sum^{n}_{i=1}   [   (\vec{k}^{~2}_{\bot    i}    +
m^{2}_{i})/x_{i}], ~~ \sum^{n}_{i=1} ~\vec{k}_{\bot i} = 0,
{}~~ \sum^{n}_{i=1}  ~x_{i}  = 1 ~~{[L.C]} \end{array}  \right.  \eqno
(3.1)
$$
from  which  one obtains, for two-particle system with  $m_{1}  =
m_{2}$ (i.e., $q^{0}_{1} = q^{0}_{2})$,
$$
\vec{q}^{~2}   \longleftrightarrow  \frac{\vec{k}^{~2}_{\bot}   +
m^{2}}{4x(1-x)} ~ - m^{2} ~~. \eqno (3.2)
$$
Then  for  two-particle  state there  is  a  possible  connection
between the rest frame wave function $\psi_{CM}(\vec{q})$,  which
controls  binding  and hadronic spectroscopy, and  the  light-cone
wave function $\psi_{LC}(x, \vec{k}_{\bot})$ by
$$
\psi_{CM}(\vec{q}^{~2})   \longleftrightarrow        \psi_{LC}
(\frac{\vec{k}^{~2}_{\bot} + m^{2}}{4x(1-x)} ~- m^{2}) ~~.  \eqno
(3.3)
$$

As an example, the wave function of the harmonic oscillator model
in  the rest frame was obtained from an approximate  bound  state
solution in the quark models for mesons$^{22}$
$$
\psi_{CM}(\vec{q}^{~2})  = A ~~  exp(-\vec{q}^{~2}/2\beta^{2})~~.
\eqno (3.4)
$$
By  using  the  connection (3.3) one  gets  the  light-cone  wave
function,
$$
\begin{array}{rl}
\psi(x_{i}, \vec{k}_{\bot}) &= A ~~exp[ - ~\frac{1}{8\beta^{2}}  (
\frac{\vec{k}^{~2}_{\bot} + m^{2}}{x_{1}} ~ + ~ \frac{\vec{k}^{~2}_{\bot}
+ m^{2}}{x_{2}} ] \\
&  = A ~~exp[ - ~\frac{\vec{k}^{~2}_{\bot} + m^{2}}{8\beta^{2} x(1-x)}  ]
{}~~. \end{array} \eqno (3.5)
$$

The  parameters  can  be  adjusted  by  using  the  first   three
constraints, i.e., Eq.(2.6)-(2.8), in Sec.II,
$$
m  = 289 ~MeV; ~~~ \beta = 385 ~MeV; ~~~ A = 0.032 ~~~  for  ~~~<
\vec{k}^{~2}_{\bot}> \approx (356 MeV)^{2} ~~.
$$
Thus  we  get  the pion  distribution  amplitude,  valence  state
structure function, and the probability $P_{q\bar{q}}$,
$$
\begin{array}{rl}
\phi(x)  &  =  \int ~(d^{2}\vec{k}_{\bot}  /  16\pi^{3})  \psi(x,
\vec{k}_{\bot}) \\
&    =    \frac{A\beta^{2}}{2\pi^{2}} x(1-x)    ~exp[    -
\frac{m^{2}}{8\beta^{2}  x(1-x)}  ]  ~~;  \end{array}  \eqno
(3.6)
$$

$$
\begin{array}{rl}
F^{V}_{2}(x)  &  =  \sum_{a=q,\bar{q}} xe^{2}_{a} ~G_{a/H}(x) \\
&    =    \frac{5A^{2}\beta^{2}}{36\pi^{2}} x^{2}(1-x)    ~exp[    -
\frac{m^{2}}{4\beta^{2} x(1-x)}  ]  ~~;  \end{array}  \eqno
(3.7)
$$

$$
\begin{array}{rl}
P_{q\bar{q}} &= \int ~(d^{2}\vec{k}_{\bot}/16\pi^{3}) \int dx \mid
\psi(x, \vec{k}_{\bot}) \mid^{2} \\
&    =    \frac{A^{2}\beta^{2}}{4\pi^{2}} \int^{1}_{0} x(1-x) ~exp[-
\frac{m^{2}}{4\beta^{2}  x(1-x)}  ] dx \\
& \approx 0.296  \end{array}  \eqno (3.8)
$$

Fig.1(a)  presents the BHL distribution amplitude  in  comparison
with that of the CZ-form and the asymptotic (AS) form$^{3}$,
$$
\phi_{AS}(x) = \sqrt{3} ~f_{\pi} x(1-x)~~, \eqno (3.9)
$$
which  is the leading term of the evolution equation of the  pion
distribution  amplitude for sufficient large $Q^{2}$. It  can  be
seen  from Fig.1(a) that the BHL distribution amplitude  is  very
close to that of the AS-form while it is significantly  different
from  that of the CZ-form. By using this  distribution  amplitude
one accounts for only $50\%$ of the pion form factor data. So  it
is  not  the "good" distribution amplitude required  to  fit  the
data.   However,   we   see  from   Fig.1(b),   where   the   NA3
parameterization   of  the  pion  structure  function   and   the
calculated  valence state structure function are presented,  that
the calculated $F^{V}_{2}(x)$ seems to be a "reasonable"  valence
state structure function. Therefore the BHL wave function  cannot
give  both  a reasonable valence state structure function  and  a
good distribution amplitude simultaneously. The fact that the BHL
wave function did not give a "good" distribution amplitude is why
the  CZ  distribution amplitude has received attention  since  it
appeared.

\vspace{0.5cm}

\noindent
{\bf B. The factorized wave functions}

In  order to fit the experimental data and to suppress  the  end-
point contributions for the applicability of perturbative QCD,  a
model  for  the pion valence wave function has been  proposed  in
Refs.12-13  by simply adding a factorized function $S(x)$ to  the
BHL wave function, with $S(x)$ specified by
$$
S(x) = (x_{1} - x_{2})^{2} = (1 - 2x)^{2} ~~, \eqno (3.10)
$$
where  $x_{1} = x$ and $x_{2} = 1 - x$. It leads  a  distribution
amplitude
$$
\phi(x)   =  \frac{A\beta^{2}}{2\pi^{2}}  ~x(1-x)(2x-1)^{2}~exp[-
\frac{m^{2}}{8\beta^{2}x(1-x)} ] ~~, \eqno (3.11)
$$
which  is of the similar shape as that of the CZ-form except  for
the  end-point regions, as shown in Fig.2(a). The parameters  are
adjusted by the first three constraints,
$$
m  =  342  ~MeV;  ~~ \beta = 455 ~MeV; ~~ A =  0.136  ~~  for  ~~
<\vec{k}^{~2}_{\bot}> \approx (343 MeV)^{2}~~.
$$
and the probability $P_{q\bar{q}}$ is 0.364, which satisfies  the
fourth  constraints, i.e., Eq.(2.9). However, the  valence  state
structure function in this case is
$$
\begin{array}{rl}
F^{V}_{2}(x) & = \sum_{a=q,\bar{q}} ~xe^{2}_{a} ~G_{a/H}(x) \\
&  = \frac{5A^{2}\beta^{2}}{36\pi^{2}} x^{2}(1-x)(2x-1)^{2}  ~exp[  -
\frac{m^{2}}{4\beta^{2} x(1-x)} ] ~~, \end{array} \eqno (3.12)
$$
which is presented in Fig.2(b). One sees from Fig.2(b) that there is
an  unreasonablely large hump in the  calculated  $F^{V}_{2}(x)$.
Thereby  the factorized wave function (3.11) though is "good"  in
giving  the CZ-like distribution which fits the data well, it  is
"bad" in giving a reasonable valence state structure function.

\vspace{0.5cm}

One may specify the factorized function $~~S(x)~~$ by other $~~$possible
distribution \\
amplitudes$^{2,3}$ which also are constrained by the
first  few  moments  given by sum rules in  terms  of  Gegenbauer
polynomials.  The distribution amplitudes in Ref.23 do  not  have
the deep dip at $x = 1/2$, hence they are different from that  of
the   CZ-form.   Fig.3  presents  the   calculated   distribution
amplitudes  and  valence state structure functions by  using  the
factorized wave functions with $S(x)$ specified by$^{23}$
$$
S(x) = 1 + 0.44 ~C^{3/2}_{2} (x_{1} - x_{2}) + 0.25 ~C^{3/2}_{4}(x_{1}
- x_{2})~~, \eqno (3.13)
$$
and
$$
S(x) = 1 + (2/3) ~C^{3/2}_{2} (x_{1} - x_{2}) + 0.43 ~C^{3/2}_{4}(x_{1}
- x_{2})~~, \eqno (3.14)
$$
respectively,   where   $C^{3/2}_{n}(\xi)$  is   the   Gegenbauer
polynomials.  The  parameters are also fixed by the  first  three
constraints in Sec.II. We see from Fig.3(a) that the distribution
amplitudes  are  more  broad  than  the  AS-form.  However,   the
calculated   valence   state  structure  functions   still   have
unreasonablely  large humps, as shown in Fig.3(b). Therefore  the
factorized  wave  functions  also  cannot  give  both  reasonable
valence   state   structure  functions  and   good   distribution
amplitudes simultaneously.

\vspace{0.5cm}

\noindent
{\bf C. The light-cone quark model wave function}

A light-cone quark model wave function for the pion was given  by
Dziembowski  and  Mankiewicz  (DM)$^{15}$  and  it  has  received
attention  for the reason that it can fit the static  properties,
the  form factor, and the CZ distribution amplitude for the  pion
simultaneously.  We indicate that this wave function,  though  is
good in "shape", has serious problems in "magnitude".

\vspace{0.5cm}

The  main idea in Ref.15 is reasonable: When one transforms  from
equal-time  (instant-form)  wave  function  to  light-cone   wave
function,  one should consider, besides the momentum  space  wave
function   transformation   such  as   the   Brodsky-Huang-Lepage
prescription, also the Melosh transformation relating  equal-time
spin  wave  functions  and light-cone spin  wave  functions.  The
following assumptions and approximations were made$^{15}$:

\begin{enumerate}
\item
It was assumed that the ground-state in the pion is described  by
the  harmonic oscillator wave function, and adopted the  Brodsky-
Huang-Lepage prescription of the momentum space wave function.
\item
They made a mock-meson assumption that mesons are a collection of
quarks  with all binding turned off and a mock-meson mass  equals
to the mean total energy of the free quarks.
\item
They adopted the mock-meson mass $M_{M} = m_{\pi}/4 + 3m_{\rho}/4
= 612~MeV$ rather than the real mass $m_{\pi} = 139 ~MeV$ or  the
mean  total  energy  of  the  two  quarks  $M_{M}  =  2<(m^{2}  +
\vec{k}^{~2})^{1/2}> ~~> ~~2(m^{2} + <\vec{k}^{~2}_{\bot}>)^{1/2}
=  950  ~MeV$ for the pion (with  $<\vec{k}^{~2}_{\bot}>  \approx
(345 MeV)^{2})$.
\item
An approximation $k^{0} + m \approx 2m$ was made in the  obtained
light-cone    wave   function,   where   $k^{0}   =   (m^{2}    +
\vec{k}^{~2})^{1/2}$.
\item
It  was impliedly assumed that the pion's helicity to be the  sum
of the light-cone helicity of quarks.
\end{enumerate}

\vspace{0.4cm}

They  got,  upon the above assumptions  and  approximations,  the
$\lambda_{1}  +  \lambda_{2}  =  0$  component  light-cone   wave
function for the pion,
$$
\psi(x,     \vec{k}_{\bot})     =    A     \frac{a_{1}a_{2}     -
\vec{k}^{~2}_{\bot}}{x(1-x)} ~exp  [  -
\frac{m^{2} + \vec{k}^{~2}_{\bot}}{8\beta^{2}x(1-x)} ] ~~,  \eqno
(3.15)
$$
where  $a_{1} = x M_{M} + m$ and $a_{2} = (1-x)M_{M} +  m$.  From
this wave function one obtains the distribution amplitude,
$$
\phi(x)              =              \frac{A}{16\pi^{2}x_{1}x_{2}}
[ a_{1}a_{2}(8\beta^{2}x_{1}x_{2})  -   (8\beta^{2}x_{1}x_{2})^{2}]
{}~exp[ - \frac{m^{2}}{8\beta^{2}x(1-x)} ] ~~. \eqno (3.16)
$$

It was argued$^{15}$ that the parameters $m = 330~MeV$ and $\beta
=  450~MeV$  are  reasonable  at  large  momentum  transfer.  The
distribution amplitude for these parameters is very close to  the
CZ-form,  as  can be seen from Fig.4. The parameter A  should  be
fixed by the first constraint in Sec.II. Then one can find,  from
(3.15), the probability of finding the $q\bar{q}(\lambda_{1} +
\lambda_{2} = 0)$ Fock state in a pion,
$$
P_{q\bar{q}} = 3.24
$$
which  is unreasonablely large. If we require that  $P_{q\bar{q}}$
be  less  than  unity,  then  the  magnitude  of  the  calculated
distribution amplitude will be less than $30\%$ of that of the CZ
distribution  amplitude, and this distribution  amplitude  surely
cannot  fit the pion form factor data. The  calculated  structure
function is,
$$
\begin{array}{rl}
F^{V}_{2}(x)  & =             \frac{5}{9}             ~~
\frac{A^{2}~x}{16\pi^{2}x_{1}^{2}x^{2}_{2}}
[2(4\beta^{2}x_{1}x_{2})^{3}                                    -
2a_{1}a_{2}(4\beta^{2}x_{1}x_{2})^{2}  \\
& + a^{2}_{1}a^{2}_{2}(4\beta^{2} x_{1}x_{2}) ]
{}~exp[ - \frac{m^{2}}{4\beta^{2}x(1-x)} ] ~~. \end{array}  \eqno
(3.17)
$$
Fig.4(b) presents the comparison of the reduced (by a factor 0.1)
valence state structure function with the NA3 parameterization of
the  pion structure function. It can be seen that the  calculated
$F^{V}_{2}$  is  almost two orders of magnitude larger  than  the
experimental  structure  function  data  as  $x  \rightarrow  1$.
Thereby  the  pion  wave  function  (3.15)  though  is  good   in
reproducing  the  shapes  of the CZ  distribution  amplitude  and
reasonable  valence  state  structure function,  it  has  serious
problems in producing the correct magnitude.

\vspace{0.5cm}

The  probability may be less than l if we change the  parameters.
For  example, we find $P_{q\bar{q}} \approx 0.86$ in the case  of
$m  = 330 ~MeV$ and $\beta = 330 ~MeV$, which were used  by  many
authors  in studying the low momentum properties for  hadrons  in
the  constituent  quark  model  framework$^{24}$.  However,   the
calculated distribution amplitude, as shown in Fig.4(a),  differs
significantly from the CZ-like distribution amplitude.

\vspace{0.6cm}

\noindent
{\bf D. The power-law wavefunction}$^{25}$

It  is interesting to look for a wave function that  can  satisfy
the four constraints:
\begin{eqnarray*}
& (1) & \int^{1}_{0}    ~dx_{1}    ~\phi(x_{i})     =
\frac{f_{\pi}}{2\sqrt{3}} \\
& (2) & \int^{1}_{0} ~dx_{1} ~\psi(x_{i}, ~\vec{k}_{\bot} = 0) =
\frac{\sqrt{3}}{f_{\pi}} \\
&  (3)  &  P_{q\bar{q}}  =  \int  ~dx_{1}  [d^{2}k_{\bot}]   \mid
\psi(x_{i}, \vec{k}_{\bot}) \mid^{2} < 1 \\
&   (4)  &  G^{(2)}_{\nu/\pi}  =  \int  ~[d^{2}\vec{k}_{\bot}  ]   \mid
\psi(x_{i}, \vec{k}_{\bot}) \mid^{2} = at ~~or ~~below ~~data.
\end{eqnarray*}
Further, we should examine the  electromagnetic
form factor $F_{\pi}(Q^{2})$ and the average transverse  momentum
$<k^{2}_{\bot}>$.

\vspace{0.5cm}

Now let us consider the power-law form of the pionic wave function
$$
\psi(x,   \vec{k}_{\bot}  =  N  ~\tilde{\phi}(\xi)   \left   (
\frac{\vec{k}^{~2}_{\bot}}{x_{1}x_{2}\beta} + 1 \right )^{-L}
$$
where  L and $\beta$ are some constants, $\xi = x_{1}  -  x_{2}$,
and the normalization
$$
N = (L - 1) \frac{16\pi^{2}}{\beta}
$$
is chosen so that
$$
\phi = x_{1} x_{2} ~\tilde{\phi}~~.
$$
For definiteness, let $\phi$ is CZ distribution amplitude, i.e.
$$
\tilde{\phi} = 5 \sqrt{3} ~f_{\pi} (x_{1}- x_{2})^{2}
$$
Then from the constraints one can get the following results:
\begin{eqnarray*}
P_{q\bar{q}}   &   =  &   \frac{9}{14}~   \frac{L-1}{2L-1}   \leq
\frac{9}{28} ~~ for ~~L \geq 1 \\
G^{(2)}_{\nu/\pi}  &  = &  \int  \frac{d^{2}k_{\bot}}{16\pi^{3}}
\mid \psi (x, \vec{k}_{\bot}) \mid^{2} = 45 \frac{L-1}{2L-1} x(1-
x)(2x-1)^{4} \\
<\vec{k}^{~2}_{\bot}>^{1/2} &=& \sqrt{\frac{40}{27}} ~\pi  ~f_{\pi}
\simeq 356 ~MeV ~~ independent ~~of ~~L
\end{eqnarray*}
In order to fit the structure function data at $x \rightarrow 1$,
we have a constraint that L is below some number. For example, we
can  put  $L  =  \frac{59}{58}$  since  a  simple  valence  quark
distribution function like
$$
G_{\nu/\pi} = \frac{3}{4} ~x^{-1/2}(1-x)
$$
gives tolerable agreement with data for $x \rightarrow 1$.

\vspace{0.5cm}

The example tell us it is possible to find wave functions that do
succeed and these are the type that should be used in any  future
calculation.

\section*{IV. The Revised Light-cone Quark Model Wave Function}

{\hskip 0.6cm}One can easily find that the assumptions 2 and 3 in
Ref.15   are  in  fact  inconsistent.  The   unreasonable   large
$P_{q\bar{q}}$  for  the  wave  function  (3.15)  should  be   an
indication  of some unreasonable assumptions  and  approximations
made  in Ref.15. Therefore we re-construct the  light-cone  quark
model  wave  function  for the pion based  upon  the  Kondratyuk-
Terent'ev  work$^{18}$  on the  relativistically  invariant  wave
function of two-particle system in light-front dynamics.

\vspace{0.5cm}

We  start  our discussion from the SU(6)  instant-form  (T)  wave
function  for the pion in the rest frame $(\vec{q}_{1} +  \vec{q}_
{2} = 0)$,
$$
\psi_{T}(\vec{q}^{~2})           =           A            ~~exp(-
\vec{q}^{~2}/2\beta^{2})(\chi^{\uparrow}_{1}\chi^{\downarrow}_{2} -
\chi^{\downarrow}_{1} \chi^{\uparrow}_{2})/\sqrt{2} ~~, \eqno (4.1)
$$
in  which  $\chi^{\uparrow, \downarrow}_{i}$  is  the  two-component
Pauli  spinor  and the two quarks have 4-momenta  $q^{\mu}_{1}  =
(q^{0},  \vec{q})$  and $q^{\mu}_{2} = (q^{0},  -\vec{q})$,  with
$q^{0}   =  (m^{2}  +  \vec{q}^{~2})^{1/2}$,  respectively.   The
instant-form spin states $\mid J, s>_{T}$ and the front-form  (F)
spin  states  $\mid  J, \lambda>_{F}$ are related  by  a  Wigner
rotation $U^{J}$ (Ref.26),
$$
\mid J, \lambda>_{F} = \sum_{s} ~U^{J}_{s\lambda} \mid J,  s>_{T}
{}~~, \eqno (4.2)
$$
and  this  rotation  is  called  Melosh  rotation  for   spin-1/2
particles.   One   should   transform   both   sides   of   (4.1)
simultaneously to get the light-cone wave function for the  pion.
For  the  left  side,  i.e.,  the  pion,  the  transformation  is
particularly  simple since the Wigner rotations are  reduced  to
unity.  For  the  right side, i.e.,  two  spin-1/2  quarks,  each
particle  instant-form and front-form spin states are related  by
the Melosh transformation$^{17-18}$,
$$
\begin{array}{rl}
\chi^{\uparrow}(T)   &   = w[(q^{+}   +   m) \chi^{\uparrow}(F)    -
q^{R}\chi^{\downarrow}(F)] ~~; \\
\chi^{\downarrow}(T)   &   = w[(q^{+}   +   m)   \chi^{\downarrow}(F) +
q^{L}\chi^{\uparrow}(F)] ~~, \end{array} \eqno (4.3)
$$
where  $w  =  [2q^{+}(q^{0}+m)]^{-1/2}, ~~ q^{R,L}  =  q^{1}  \pm
iq^{2}$, and $q^{+} = q^{0} + q^{3}$. We also adopt the  Brodsky-
Huang-Lepage  prescription for the harmonic  oscillator  momentum
space  wave function transformation. Then we get  the  light-cone
(or front-form) wave function for the pion,
$$
\psi(x,         \vec{k}_{\bot})        =         A         ~exp[-
\frac{m^{2}+\vec{k}^{~2}_{\bot}}{8\beta^{2}x(1-x)} ]
\sum_{\lambda_{1},  \lambda_{2}}  ~C^{F}_{0}(x,   \vec{k}_{\bot},
\lambda_{1},         \lambda_{2})    \chi^{\lambda_{1}}_{1}(F)
\chi^{\lambda_{2}}_{2}(F)~~, \eqno (4.4)
$$
where  the component coefficients  $C^{F}_{0}(x,  \vec{k}_{\bot},
\lambda_{1}, \lambda_{2})$ for J = 1, when expressed in terms  of
the instant-form momentum $q^{\mu} = (q^{0}, \vec{q})$, have  the
forms,
$$
\begin{array}{rl}
C^{F}_{0}(x,    \vec{k}_{\bot},    \uparrow,    \downarrow) &=
w_{1}w_{2}[(q^{+}_{1}  + m)(q^{+}_{2} + m) -  \vec{q}^{~2}_{\bot}]
/\sqrt{2} ~~; \\
C^{F}_{0}(x,    \vec{k}_{\bot},   \downarrow, \uparrow)  &=
-w_{1}w_{2}[(q^{+}_{1}  + m)(q^{+}_{2} + m) -  \vec{q}^{~2}_{\bot}]
/\sqrt{2} ~~; \\
C^{F}_{0}(x,    \vec{k}_{\bot},    \uparrow, \uparrow)  &=
w_{1}w_{2}[(q^{+}_{1}  + m)q^{L}_{2}-(q^{+}_{2} + m)
q^{L}_{1}]/\sqrt{2} ~~; \\
C^{F}_{0}(x,    \vec{k}_{\bot},    \downarrow, \downarrow)  &=
w_{1}w_{2}[(q^{+}_{1}  + m)q^{R}_{2}-(q^{+}_{2} + m)
q^{R}_{1}]/\sqrt{2} ~~; \end{array} \eqno (4.5)
$$
which satisfy the relation,
$$
\sum_{\lambda_{1},  \lambda_{2}}  ~C^{F}_{0}(x,   \vec{k}_{\bot},
\lambda_{1},   \lambda_{2})^{*}   ~C^{F}_{0}(x,   \vec{k}_{\bot},
\lambda_{1}, \lambda_{2}) = 1 ~~. \eqno (4.6)
$$
It can be seen that there are two higher helicity $(\lambda_{1} +
\lambda_{2} = \pm 1)$ components in the expression of the  light-
cone  wave  function for the pion besides the  ordinary  helicity
$(\lambda_{1}  + \lambda_{2} = 0)$ components. These  two  higher
helicity  components  arise  from Wigner  rotations  (or  Melosh
rotations  in a strict sense)$^{27}$. We also indicate  that  one
should  express  the  instant-form momentum  $\vec{q}  =  (q^{3},
\vec{q}_{\bot})$   in   terms   of   the   light-cone    momentum
$\underline{k}   =  (x,  \vec{k}_{\bot})$  in  the   calculation.
However, the relation between $\vec{q}$ and $\underline{k}$ is by
no  means unique, thus in practice one needs to construct  models
relating  them.  This  leads  us to  discuss  the  following  two
possible  schemes relating the instant-form momentum  $\vec{q}  =
(q^{3},    \vec{q}_{\bot})$   and   the    light-cone    momentum
$\underline{k} = (x, \vec{k}_{\bot})$.

\vspace{0.5cm}

\noindent
{\bf a). Scheme one}

A reasonable connection between $\vec{q}$ and $\underline{k}$ was
given in Ref.28-29,
$$
\begin{array}{rl}
x & = (q^{0} + q^{3})/M~~; \\
\vec{k}_{\bot} & = \vec{q}_{\bot} ~~, \end{array} \eqno (4.7)
$$
in which M satisfies
$$
M^{2}  =  \frac{\vec{k}^{~2}_{\bot} +  m^{2}}{x(1-x)}  ~~.  \eqno
(4.8)
$$
{}From Eq.(4.7) we find,
$$
\begin{array}{rl}
q^{0} & = [xM + (m^{2} + \vec{k}^{~2}_{\bot})/xM]/2 ~~; \\
q^{3}   &  =  [xM  -  (m^{2}  +  \vec{k}^{~2}_{\bot})/xM]/2   ~~,
\end{array} \eqno (4.9)
$$
thus
$$
\begin{array}{rl}
& q^{+} = xM ~~, \\
&  2q^{+}  (q^{0} + m) = (xM + m)^{2} +  \vec{k}^{~2}_{\bot}  ~~.
\end{array} \eqno (4.10)
$$
We notice
$$
\vec{q}^{~2}  = \frac{\vec{k}^{~2}_{\bot} + m^{2}}{4x(1-x)}  ~  -
m^{2} ~~, \eqno (4.11)
$$
which  is  consistent with Eq.(3.2) in  the  Brodsky-Huang-lepage
prescription$^{3}$. The detailed reasons for the connection (4.7)
can  be found in Refs.28-29. Thus the $\lambda_{1}+\lambda_{2}  =
0$ component wave function can be obtained,
$$
\psi(x,     \vec{k}_{\bot})     =     A     \frac{a_{1}a_{2}  -
\vec{k}^{~2}_{\bot}}{[(a_{1}^{2}                                +
\vec{k}^{~2}_{\bot})(a^{2}_{2} +  \vec{k}^{~2}_{\bot})]^{1/2}}
{}~exp[  - \frac{m^{2} +  \vec{k}^{~2}_{\bot}}{8\beta^{2}x(1-x)}  ]
{}~~, \eqno (4.12)
$$
where  $a_{1} = xM + m$ and $a_{2} = (1 -x)M +m$. The  constraint
$P_{q\bar{q}}  < 1$ will be satisfied with any reasonable  m  and
$\beta$  by  using the first constraint to fix the  parameter  A.
Fig.5   presents  the  calculated  distribution  amplitudes   and
structure  functions with two sets of parameters, i.e., $m =  330
{}~MeV,  ~~  \beta  =  540  ~MeV$, and $m  =  \beta  =  330  ~MeV$,
respectively,  for the wave function (4.12). When  compared  with
the results from the BHL wave function (3.15) in Fig.1, we  find,
in contrary to the claims in Ref.15, that the effect from  Melosh
rotation  seems  to  have  no  significant  effect  on  both  the
calculated distribution amplitude and the valence state structure
function.  The  corresponding  $P_{q\bar{q}}$ for  the  two  set
parameters are 0.228 and 0.552, respectively.

\vspace{0.5cm}

\noindent
{\bf b). Scheme two}

We adopt the assumptions 1 and 2 in Sec.III as the starting point
for  scheme two. In this case the relation between $\vec{q}$  and
$\underline{k}$ is very simple,
$$
\begin{array}{rl}
& x = (q^{0} + q^{3})/M ~~; \\
& \vec{k}_{\bot} = \vec{q}_{\bot} ~~, \end{array} \eqno (4.13)
$$
where M, the mock-meson mass, defined to be the mean total energy
of the free quarks,
$$
M  =  2 < (m^{2} + \vec{q}^{~2})^{1/2} > \approx  2(m^{2}  +  3/2
<\vec{k}^{~2}_{\bot}>)^{1/2} ~~, \eqno (4.14)
$$
the value of which should be, approximately, 1130 MeV and 910 MeV
for the above two sets of parameters with $<\vec{k}^{~2}_{\bot}>$
being  $(374  ~MeV)^{2}$  and $(256  ~MeV)^{2}$  respectively  in
scheme one. From Eq.(4.13) it follows,
$$
\begin{array}{rl}
& q^{+} = xM ~~; \\
& 2q^{+}(q^{0} + m) = (xM + m)^{2} + \vec{k}^{~2}_{\bot} ~~,
\end{array}
$$
Thereby  the  $\lambda_{1}  + \lambda_{2}  =  0$  component  wave
function should be (4.12) with fixed M rather than M in Eq.(4.8).
We find $P_{q\bar{q}}$ to be 0.484 and 0.723 respectively,  which
are larger than those in scheme one. The calculated  distribution
amplitudes  and structure functions, as presented in  Fig.6,  are
also  larger  than  those in scheme one. We  also  calculate  the
distribution  amplitude and the valence state structure  function
for  the first set parameters by using M = 612 MeV  and  compared
them with the results in Ref.15 as shown in Fig.4. It can be seen
that  the  distribution amplitude is close to the  CZ-form,  with
also the unreasonablely large valence state structure function as
that in Sec.III.C. The probability $P_{q\bar{q}}$ is 2.93,  which
is  larger  than  unity.  Thus the flaws  suffered  by  the  wave
function (3.15) mainly raised from the inconsistent assumption 3;
i.e.,  the  use of a smaller M. The approximation 4  also  has  a
large  consequence  on the results,  thereby  this  approximation
seems to be too strong. We also notice that
$$
\vec{q}^{~2}    =   \frac{1}{4}   ~[xM   +   (   \frac{m^{2}    +
\vec{k}^{~2}_{\bot}}{xM})]^{2} - m^{2} ~~, \eqno (4.16)
$$
which  is inconsistent with Eq.(3.2). Further observation of  the
unreasonableness of the results in Ref.15 is that the effect from
Melosh  rotation should disappear for $\vec{k}_{\bot} = 0$.  This
aspects is satisfied for the wave function (4.12), whereas it  is
not satisfied for the wave function (3.15).

\vspace{0.5cm}

We only comment without argument here that there are  ambiguities
of  introducing a factor, such as $\sqrt{1/2x_{1}x_{2}}$  adopted
in  Ref.28  or $\sqrt{M/4x_{1}x_{2}}$ adopted in Ref.29,  to  the
Brodsky-Huang-Lepage  wave  function as a  consequence  from  the
jacobian relating instant-form momentum and light-cone  momentum.
However,  the qualitative conclusions in above analyses will  not
changed though the quantitative results will be different if  the
factor is introduced.

\vspace{0.5cm}

\section*{V. Summary and Comments}

{\hskip  0.6cm}After the analysis of several existing  pion  wave
functions in light-cone formalism and the re-construction of  the
light-cone  quark  model  wave function, it  is  shown  that  the
present   wave  functions  $(\lambda_{1}  +  \lambda_{2}   =   0$
component)  have  failed the requirements to fit  the  pion  form
factor  data  and  the reasonable valence  Fock  state  structure
function  which does not exceed the pion structure function  data
with reasonable parameters. However, as an example, we consider a
power-law form of the pionic wave function. It is shown that  all
of  constraints  can be satisfied as $L  =  \frac{59}{58}$.  This
means that it is possible to find wave functions that do  succeed
and  these  are  the  type that should  be  used  in  any  future
calculation.

\vspace{0.5cm}

We  also  find, in contrary to pervious claims, that  the  effect
from  Melosh  rotation in the ordinary helicity  $(\lambda_{1}  +
\lambda_{2} = 0)$ component wave function cannot reproduce a  CZ-
like  distribution amplitude. However, as mentioned in Sec.I,  it
is  still  an  open  problem  whether  the  CZ-like  distribution
amplitude  is  the correct pion distribution amplitude,  and  the
large  second  moment of the CZ distribution  amplitude  are  not
reproduced by a recent improved lattice QCD calculation. Thus  we
could  consider  the quark model evaluation of this  paper  as  a
suggestion that the pion distribution amplitude may close to  the
asymptotic form rather than to the CZ form.

\vspace{0.5cm}

Then  we  meet  the  problem that  the  "naive"  asymptotic  pion
distribution  amplitude can only account for about $50\%$ of  the
existing pion form factor data. In previous perturbative QCD work
the adoption of a CZ-like distribution amplitude was found to  be
a  possible  way to resolve this problem. We  speculate  in  this
paper  that  the introduction of the higher helicity  states  may
provide an alternative way to address the problem concerning  the
applicability   of  perturbative  QCD.  These   higher   helicity
components emerge naturally in the full light-cone wave  function
as  a consequence from the Wigner rotation$^{26}$ relating  spin
states in different frames. The existence of the higher  helicity
components in the light-cone wave function for the pion is not in
contradiction   with   the  requirement   of   angular   momentum
conservation,  since  the full light-cone  wave  function,  e.g.,
(4.4),  is  an eigenstate of the total spin operator  $I^{2}$  in
front-form dynamics$^{29}$.

\vspace{0.5cm}

The introduction of the higher helicity component wave  functions
into the light-cone formalism may have significant consequence in
several  problems  concerning the applicability  of  perturbative
QCD.  It is speculated that the perturbative  contributions  from
the higher helicity states may provide the other fraction  needed
to  fit  the  pion  form factor  data  besides  the  perturbative
contributions  from  the ordinary helicity  states  evaluated  by
using  the "naive" Brodsky-Huang-Lepage  distribution  amplitude.
However,  there are still some difficulties in incorporating  the
higher  helicity component wave functions into  the  conventional
perturbative QCD framework since the distribution amplitudes  for
these components vanish from Eq.(2.5). Thereby it is necessary to
developed   a  technique  which  enables  us  to  calculate   the
perturbative  contributions from these components. Some  progress
has been made in this direction and it will be given elsewhere.

\baselineskip 20pt
\newpage
\section*{Figure Captions}
\begin{description}
\item[Fig.1](a). The   normalized    distribution    amplitude
$\hat{\phi}(x)  = \phi(x)/\sqrt{3} ~f_{\pi}$: curves CZ and  AS
are  the Chernyak-Zhitnitsky distribution amplitude  (Ref.4)  and
the  asymptonic distribution amplitude (Ref.3); curve BHL is  the
distribution   amplitude  from  the   Brodsky-Huang-Lepage   wave
function   (3.5)  (Ref.3),  respectively.  (b).   The   structure
function:  curve  NA3  is the NA3 parameterization  of  the  pion
structure  function;  curve BHL is the  valence  state  structure
function for the Brodsky-Huang-Lepage wave function.
\item[Fig.2]Similar as in Fig.1. (a). Curve H is the distribution
amplitude  (3.11)  from Huang's factorized  ansatz  (Refs.12-13).
(b). Curve H is the valence state structure function for  Huang's
factorized wave function.
\item[Fig.3] Similar as in Fig.1. Curves BF(I) and BF(II) are the
results  from  the  factorized wave functions,  with  $S(x_{1}  -
x_{2})$   to  be  the  two  sets  Braun-Filyanov  (Ref.24)   like
distribution  amplitudes,  for the parameters $m = 324  ~MeV,  ~~
\beta  =  432  ~MeV$, and $m = 346 ~MeV, ~~  \beta  =  461  ~MeV$,
respectively.
\item[Fig.4]Similar  as in Fig.1. Curves DM(1) and DM(2) are  the
results  from the Dziembowski-Mankiewicz (Ref.15)  wave  function
(3.15)  with  the parameters to be $m = 330 ~MeV, ~~\beta  =  450
{}~MeV$, and $m = 330 ~MeV, ~~\beta = 330 ~MeV$, respectively.
\item[Fig.5]Similar  as in Fig.1. Curves a and b are the  results
from  the $(\lambda_{1} + \lambda_{2} = 0)$ component  light-cone
wave function (4.12), with M in Eq.(4.8), for scheme one with the
parameters  to be $m = 330~MeV, ~~\beta = 540~MeV$, and $m =  330
{}~MeV, ~~ \beta = 330 ~MeV$, respectively.
\item[Fig.6]  Similar  as  in Fig.1. Curves a, b and  c  are  the
results  from  the $(\lambda_{1} + \lambda_{2}  =  0)$  component
light-cone  wave  function (4.12), with fixed M, for  scheme  two
with  the parameters to be: $m = 330 ~MeV, ~~ M = 1130 ~MeV$  and
$\beta = 540 ~MeV; ~~ m = 330 ~MeV, ~~ M = 910 ~MeV$ and $\beta =
330  ~MeV$;  and $m = 330 ~MeV, M = 612 ~MeV$ and  $\beta  =  540
{}~MeV$.
\end{description}

\newpage
\begin{thebibliography}{99}
\bibitem{1}See,  e.g.,  S.  J.  Brodsky  and  G.  P.  Lepage,  in
Perturbative  Quantum  Chromodynamics, edited by  A.  H.  Mueller
(World Scientific, Singapore, 1989); and references therein.
\bibitem{2}G.  P.  Lepage  and S. J.  Brodsky,  Phys.  Rev.  D22,
2157(1980);  S.  J. Brodsky, Y. Frishman, G. P.  Lepage,  and  C.
Sachrajda, Phys. Lett. 91B, 239(1980).
\bibitem{3}S.  J.  Brodsky,  T.  Huang,  and  G.  P.  lepage,  in
Particles  and  Fields,  edited by A. Z. Capri and  A.  N.  Kamal
(Plenum  Publishing  Corporation, New York, 1983), p.143;  G.  P.
Lepage, S. J. Brodsky, T. Huang and P. B. Mackenzie, ibid,  p.83;
T.  Huang,  in Proceedings of XX-th International  Conference  on
High Energy Physics, Madison, Wisconsin, July 17-23, 1980, edited
by  L. Durand and L. G. Pondrom (American Institute  of  Physics,
New York, 1981), p.1000.
\bibitem{4}V. L. Chernyak and A. R. Zhitnitsky, Nucl. Phys. B201,
492(1982); Phys. Rep. 112, 173(1984); Nucl. Phys. B246, 52(1984).
\bibitem{5}T.  Huang, X. D. Xiang, and X. N. Wang, Chinese  Phys.
Lett. 2, 76(1985); Commun. Theor. Phys. 5, 117(1986); Phys.  Rev.
D35, 1013(1987).
\bibitem{6}S.  Gottlieb and A. S. Kronfeld, Phys. Rev. Lett.  55,
2531(1985);  Phys. Rev. D33, 227(1986); G. Martinelli and  C.  T.
Sachrajda, Phys. Lett. 217B, 319(1989).
\bibitem{7}D.  Daniel, R. Gupta, and D. G. Richards,  Phys.  Rev.
D43, 3715(1991).
\bibitem{8}E.  E.  Salpeter  and  H. A.  Bethe,  Phys.  Rev.  84,
1232(1951).
\bibitem{9}H.  C. Pauli and S. J. Brodsky, Phys. Rev. D32,  1993,
2001(1985); T. Eller, H. C. Pauli, and S. J. Brodsky, Phys.  Rev.
D35,  1493(1987). See, also, A. C.  Tang,  SLAC-Report-351(1990),
and references therein.
\bibitem{10}S.  J. Brodsky, C.-R. Ji, and M. Sawicki, Phys.  Rev.
D32, 1530(1985); T. Huang and Z. Huang, Commun. Theor. Phys.  11,
479(1989).
\bibitem{11}O.  C. Jacob and L. S. Kisslinger, Phys. Lett.  243B,
323(1990);  L.  S.  Kisslinger and O. C. Jacob,  in  Nuclear  and
Particle  Physics on the Light Cone, edited by M. B. Johnson  and
L. S. Kisslinger (World Scientific, Singapore, 1989), p.322.
\bibitem{12}T.   Huang,  in  Proceedings  of  the   International
Symposium  on Particle and Nuclear Physics, Beijing,  Sept.  2-7,
1985, edited by N. Hu and C. S. Wu, (World Scientific, Singapore,
1986),  p.151; and in Proceedings of Socond Asia Pacific  Physics
Conference,  India,  1986, (World Scientific,  Singapore,  1987),
p.258.
\bibitem{13}T.  Huang, Nucl. Phys. B(Proc. Suppl.)7B,  320(1989);
T. Huang and Q. X. Shen, Z. Phys. C50, 139(1991).
\bibitem{14}C.   E.  Carson  and  F.  Gross,  Phys.   Rev.   D36,
2060(1987).
\bibitem{15}Z.  Dziembowski and L. Mankiewicz, Phys.  Rev.  Lett.
58, 2175(1987); Z. Dziembowski, Phys. Rev. D37, 778(1987); and in
Nuclear  and Particle Physics on the Light Cone, edited by M.  B.
Johnson and L. S. Kisslinger (World Scientific, Singapore, 1989),
p.166.
\bibitem{16}See,  e.g.,  W. J. Metcalf et al., Phys.  Lett.  91B,
275(1980).
\bibitem{17}H. J. Melosh, Phys. Rev. D9, 1095(1974).
\bibitem{18}L.  A. Kondratyuk and M. V. Terent'ev, Yad. Fiz.  31,
1087(1980) [Sov. J. Nucl. Phys. 31, 561(1980)].
\bibitem{19}J. Badier et al., Z. Phys. C18, 281(1983).
\bibitem{20}A.  Altarelli  and  G.  Parisi,  Nucl.  Phys.   B126,
298(1977).
\bibitem{21}B. Q.  Ma, Phys. Lett. 176B, 179(1986); B.Q. Ma  and
J. Sun, Int. J. Mod. Phys. A6, 345(1991).
\bibitem{22}See,  e.g., Elementary Particle Theory Group,  Peking
University,  Acta Physics Sinica 25, 415(1976); N. Isgur, in  The
New Aspects of Subnuclear Physics, edited by A. Zichichi (Plenum,
New York, 1980), p.107.
\bibitem{23}V.  M.  Braun  and  I. E.  Filyanov,  Z.  Phys.  C44,
157(1989).
\bibitem{24}H.  J.  Weber, Ann. Phys. (N.  Y.)177,  38(1987);  K.
Konen and H. J. Weber, Phys. Rev. D41, 2201(1990), and references
therein.
\bibitem{25}Thanks referee for making a comment on the  power-law
wave function.
\bibitem{26}E. Wigner, Ann. Math. 40, 149(1939).
\bibitem{27}N. Isgur and C. H. Llewellyn Smith, Nucl. Phys. B317,
526(1989).
\bibitem{28}V.  A. Karmanov, Nucl. Phys. B166,  378(1980);  ibid.
A362, 331(1981).
\bibitem{29}F.  Coester, in Constraint's Theory and  Relativistic
Dynamics,  edited by G. Longhi and L. Lusanna (World  Scientific,
Singapore,  1987),  p.159;  and in The Three-Body  Force  in  the
Three-Nucleon  System,  edited by B. L. Berman and B.  F.  Gibson
(Springer, New York, 1986), p.472.
\end{thebibliography}
\end{document}